\documentclass[letterpaper]{jpconf}
\usepackage{ae}
\usepackage{bm}
\usepackage{color}
\usepackage{amsmath}
\usepackage{amssymb}
\usepackage{amsfonts}
\usepackage{graphicx}
\usepackage{slashed}
\usepackage{wasysym}
\usepackage{setspace}
\usepackage{cite}

\newcommand{\Eqref}[1]{Eq.~\eqref{#1}}

\renewcommand{\k}{\hat{\vec{k}}}
\newcommand{\rk}{{\rm k}}

\newcommand{\f}{\varphi}
\renewcommand{\t}{\vartheta}

\renewcommand{\vec}[1]{\mathbf{#1}}

\begin{document}

\title{Vacuum Birefringence as a Vacuum Emission Process}

\author{Felix Karbstein}

\address{Helmholtz-Institut Jena
\& Theoretisch-Physikalisches Institut, Abbe Center of Photonics, \\ Friedrich-Schiller-Universit\"at Jena, Max-Wien-Platz 1, 07743 Jena, Germany}

\ead{felix.karbstein@uni-jena.de}

\begin{abstract}
 We argue that the phenomenon of vacuum birefringence in strong inhomogeneous electromagnetic fields can be most efficiently analyzed in terms of a vacuum emission process.
 In this contribution, we exemplarily stick to the case of vacuum birefringence in a stationary perpendicularly directed, purely magnetic background field extending over a finite spatial extent. Similar field configurations are realized in the BMV and PVLAS experiments.
 We demonstrate that we can reproduce the conventional constant field result. Our focus is on effects which arise when the probe photons originate in the field free region, are directed towards the magnetic field region, and detected well after the interaction with the magnetic field has taken place, again at zero field. 
\end{abstract}

\section{Introduction}

Virtual charged particle fluctuations in the quantum vacuum
give rise to nonlinear, effective couplings between electromagnetic fields \cite{Euler:1935zz,Heisenberg:1935qt,Weisskopf}.
Vacuum birefringence in strong electromagnetic fields constitutes one of the most prominent optical signatures of quantum vacuum nonlinearity arising within the framework of quantum electrodynamics (QED) \cite{Toll:1952,Baier,BialynickaBirula:1970vy,Adler:1971wn,Kotkin:1996nf}. It is so far searched for in experiments using macroscopic magnetic fields \cite{Cantatore:2008zz,Berceau:2011zz}.

An alternative proposal to verify vacuum birefringence with the aid of high-intensity lasers in an all-optical experimental setup has been put forward by \cite{Heinzl:2006xc},
envisioning the combination of an optical high-intensity laser as pump and a linearly polarized x-ray pulse as probe.
Resorting to the locally constant field approximation on the level of the effective Lagrangian, recently we have reanalyzed this vacuum birefringence scenario \cite{Karbstein:2015xra}, rephrasing the phenomenon in terms of a vacuum emission process \cite{Karbstein:2014fva}.

Here, we adopt the same approach to the case of vacuum birefringence in a stationary, perpendicularly directed, purely magnetic background field extending over a finite spatial extent, and show that we can recover the conventional constant field result.
While this rederivation serves as a simple illustration of our method, we also point out deviations arising beyond infinitely extended constant background fields:
Apart from vacuum birefringence, probe photons originating at zero field, that are directed towards a magnetic field inhomogeneity can experience the phenomenon of quantum reflection \cite{Gies:2013yxa,Gies:2014wsa}. Of course, magnetic fields with a finite extent only constitute corresponding magnetic field inhomogeneities.

For other experimental signatures of quantum vacuum nonlinearities, we refer to the pertinent reviews about this topical research field \cite{Dittrich:2000zu,Marklund:2008gj,Dunne:2008kc,Heinzl:2008an,DiPiazza:2011tq,Battesti:2012hf} and references therein.

\section{The locally constant field approximation}

For constant electromagnetic fields, the effective Lagrangian encoding quantum corrections to the Maxwell Lagrangian of classical electrodynamics is a function of the  gauge and Lorentz invariants of the electromagnetic field
${\cal F}=\frac{1}{4}F_{\mu\nu}F^{\mu\nu}=\frac{1}{2}(\vec{B}^2-\vec{E}^2)$ and ${\cal G}=\frac{1}{4}F_{\mu\nu}{}^*F^{\mu\nu}=-\vec{E}\cdot\vec{B}$ only;
$^*F^{\mu\nu}=\frac{1}{2}\epsilon^{\mu\nu\alpha\beta}F_{\alpha\beta}$ is the dual field strength tensor, with $\epsilon^{0123}=1$.
Our metric convention is $g_{\mu \nu}=\mathrm{diag}(-1,+1,+1,+1)$, and we use the Heaviside-Lorentz System with $c=\hbar=1$.
Moreover, Furry's theorem (charge conjugation symmetry of QED) demands the effective Lagrangian to be even in the number of couplings to the electromagnetic field.

The electron mass $m$ is the only other physical energy scale available in QED in constant electromagnetic fields and in the absence of real electrons and photons.
Hence, for dimensional reasons, the quantum corrections to the Maxwell Lagrangian vanishing in the limit $\hbar\to0$ are of the following structure,
\begin{equation}
 {\cal L}=m^4f\bigl(\tfrac{\cal F}{m^4},\tfrac{\cal G}{m^4}\bigr) \quad\leftrightarrow\quad
 {\cal L}=\Bigl({\cal F}\frac{\partial}{\partial{\cal F}}+{\cal G}\frac{\partial}{\partial{\cal G}}+m^4\frac{\partial}{\partial m^4}\Bigr){\cal L}\,. \label{eq:Lgen}
\end{equation}
At one-loop level, this Lagrangian amounts to the renowned Heisenberg-Euler effective Lagrangian \cite{Heisenberg:1935qt}, whose low-field expansion reads
\begin{equation}
 \frac{{\cal L}}{m^4}=\frac{1}{4\pi^2}\frac{1}{90}\Bigl(\frac{e}{m^2}\Bigr)^4\Bigl[(4{\cal F}^2+7{\cal G}^2)-\frac{4}{7}\Bigl(\frac{e}{m^2}\Bigr)^2{\cal F}(8{\cal F}^2+13{\cal G}^2)\Bigr]+{\cal O}\bigl((\tfrac{e\epsilon}{m^2})^8\bigr), \label{eq:L_HE}
\end{equation}
where we have counted $F^{\mu\nu}$ and ${}^*F^{\mu\nu}$ as ${\cal O}(\varepsilon)$.

In inhomogeneous backgrounds additional gauge and Lorentz invariant building blocks become available.
For slowly varying fields the deviations from the constant field limit can be accounted for with derivative terms $\sim\partial_\alpha F^{\mu\nu}$.
If the typical frequency/momentum scale of variation of the inhomogeneous background field is $\upsilon$, which can be related to a typical time/length scale of variation $d$
via $\upsilon\sim\frac{1}{d}$, derivatives effectively translate into multiplications with $\upsilon$ to be rendered dimensionless by the electron mass $m$.
Adopting the constant field result~\eqref{eq:Lgen} for slowly varying inhomogeneous fields by means of the locally constant field approximation, the deviations from the exact result are of ${\cal O}\bigl((\tfrac{\upsilon}{m})^2\bigr)$ \cite{Karbstein:2015cpa}.
Employing the locally constant field approximation on the level of the constant-field Lagrangian~\eqref{eq:Lgen} allows for trustworthy results as long as $d\gg\frac{1}{m}$, i.e.,
the typical time/length scales governing the variation of the field inhomogeneity have to be significantly larger than the Compton time $\tau_C$/Compton wavelength $\lambda_C$ of the electron;
$\tau_c=\frac{1}{m}\approx1.3\cdot10^{-21}\,{\rm s}$ and $\lambda_C=\frac{1}{m}\approx3.9\cdot10^{-13}\,{\rm m}$.
Hence, this approximation is certainly well justified for the typical field configurations accessible in the laboratory.

\section{Stimulated single-photon emission from the vacuum}

Here we aim at analyzing vacuum birefringence in terms of a vacuum emission process.
As detailed in \cite{Karbstein:2014fva}, the single photon emission amplitude from the vacuum subjected to slowly varying electromagnetic fields is given by
\begin{equation}
 {\cal S}_{(p)}(\vec{k})\equiv\langle\gamma_{p}(\vec{k})|\int{\rm d}^4x\, f^{\mu\nu}(x)\frac{\partial{\cal L}}{\partial F^{\mu\nu}}(x)|0\rangle\,, \label{eq:Sp1}
\end{equation}
with the single photon state denoted by $|\gamma_{p}(\vec{k})\rangle\equiv a^\dag_{\vec{k},p}|0\rangle$. Here $p\in\{1,2\}$ denotes the polarization of the emitted photon of four wave-vector $k^\mu=(\rk,\vec{k})$ and $\rk\equiv|\vec{k}|$; the unit wave-vector is $\k=\vec{k}/\rk$.
Equation~\eqref{eq:Sp1} can be recast into
\begin{multline}
{\cal S}_{(p)}(\vec{k})
 =\frac{i}{\sqrt{2\rk}} \hat f^{\mu\nu}_{(p)}(k)\int{\rm d}^4x\,{\rm e}^{ikx}\frac{\partial{\cal L}}{\partial F^{\mu\nu}}(x) \\
 =\frac{1}{45}\frac{m^2}{8\pi^2}\frac{i e}{\sqrt{2\rk}}\biggl\{ \Bigl(\frac{e}{m^2}\Bigr)^3\!\int{\rm d}^4x\,{\rm e}^{ikx} \Bigl[4{\cal F}(x)F_{\mu\nu}(x) +7{\cal G}(x){}^*F_{\mu\nu}(x) \Bigr]\hat f^{\mu\nu}_{(p)}(k) + {\cal O}\bigl((\tfrac{e\varepsilon}{m^2})^5\bigr)\biggr\}\,,
 \label{eq:Sp2}
\end{multline}
where $\hat f^{\mu\nu}_{(p)}(k)=k^\mu\epsilon^{*\nu}_{(p)}(k)- k^\nu\epsilon^{*\mu}_{(p)}(k)$ denotes the normalized field strength tensor of the emitted photon in momentum space, and we have inserted \Eqref{eq:L_HE} in the last step.

In spherical coordinates, we have $\k=(\cos\f\sin\t,\sin\f\sin\t,\cos\t)$, and the two transversal photon polarization modes can be spanned by the
two linear polarization vectors
\begin{equation}
 \epsilon^\mu_{(1)}(\k)=(0,\vec{e}_{\perp,\chi}) \quad\textrm{and}\quad \epsilon^\mu_{(2)}(\k)=\epsilon^\mu_{(1)}(\k)\big|_{\chi\to\chi+\frac{\pi}{2}}\,, \label{eq:epsilons}
\end{equation}
where
\begin{equation}
\vec{e}_{\perp,\chi}=
\left(\begin{array}{c}
  \cos\f\cos\t\cos\chi-\sin\f\sin\chi \\
  \sin\f\cos\t\cos\chi+\cos\f\sin\chi \\
  -\sin\t\cos\chi
 \end{array}\right), \label{eq:epsilons1}
\end{equation}
fulfilling $\k\cdot\vec{e}_{\perp,\chi}=0$.
Important constituents of \Eqref{eq:Sp2} are the contractions $F_{\mu\nu}(x)\hat f^{\mu\nu}_{(p)}(k)$ and ${}^*F_{\mu\nu}(x)\hat f^{\mu\nu}_{(p)}(k)$.
Because of the antisymmetry of the field strength tensors in the indices $\mu$ and $\nu$, it suffices to specify just six independent components,
which we collect in the following table
\begin{equation}
 \begin{array}{c||c|c|c|}
 \mu\nu & \hat f^{\mu\nu}_{(1)}(k) & F_{\mu\nu} & {}^*F_{\mu\nu} \\
 \hline\hline
 10 & \rk(\sin\f\sin\chi - \cos\f\cos\t\cos\chi) & E_1 & B_1 \\
 20 & - \rk(\sin\f\cos\t\cos\chi+\cos\f\sin\chi) & E_2 & B_2 \\
 30 & \rk \sin\t\cos\chi & E_3 & B_3 \\
 12 & \rk\sin\t\sin\chi & B_3 & -E_3 \\
 13 & -\rk(\cos\f\cos\chi-\cos\t\sin\f\sin\chi) & -B_2 & E_2 \\
 23 & -\rk(\sin\f\cos\chi+\cos\t\cos\f\sin\chi) & B_1 & -E_1 \\
 \hline
 \end{array}
\end{equation}
for further reference. The above results are very generic and allow for the determination of single photon emission signals from any field profile compatible with the locally constant field approximation.

\section{Vacuum birefringence as a vacuum emission process}

Aiming at the study of vacuum birefringence in terms of a vacuum emission process from a given macroscopic field configuration,
the field configuration inducing the outgoing photon signal is the superposition of both the propagating incident probe photon field
and the external electromagnetic field ``polarizing'' the quantum vacuum. 

Here, we exemplarily focus on a linearly polarized plane-wave probe traversing a perpendicularly directed magnetic field.
We assume the direction of the magnetic field to be globally fixed, and to point into $\rm z$ direction, i.e., $\vec{B}_0=B\vec{e}_{\rm z}$.
The probe photons propagate along $\rm x$.
We parameterize the magnetic and electric fields of the probe photon wave as $\vec{b}=b\,\vec{e}_\beta$ and $\vec{e}=b\,\vec{e}_{\beta+\frac{\pi}{2}}$, with $\vec{e}_\beta\equiv(0,\sin\beta,\cos\beta)$; the choice of the angle $\beta$ fixes the polarization vector of the probe field.
More specifically, $b\equiv b(x)=b_0\cos(\omega(t-{\rm x}))$, with probe field amplitude $b_0$ and energy $\omega>0$.
Of course, our approach could be straightforwardly generalized to more general probes beyond the plane-wave limit also.
Correspondingly, the superposition of the pump and probe fields results in the macroscopic electromagnetic fields
$\vec{B}=B\vec{e}_{\rm z}+b\,\vec{e}_\beta$ and $\vec{E}=b\,\vec{e}_{\beta+\frac{\pi}{2}}$, for which we obtain
\begin{gather}
 {\cal F} = \frac{1}{2}B^2+B b \cos\beta \,, \quad {\cal G} = Bb\sin\beta \,, \\
 F_{\mu\nu}\hat f^{\mu\nu}_{(1)}(k) = 2{\rm k}\bigl\{b\bigl[(\sin\t-\cos\f)\sin(\chi-\beta) - \sin\f\cos\t \cos(\chi-\beta)\bigr] + B\sin\t\sin\chi\bigr\} \,, \\
 {}^*F_{\mu\nu}\hat f^{\mu\nu}_{(1)}(k) = F_{\mu\nu}\hat f^{\mu\nu}_{(1)}(k)\big|_{\chi\to\chi+\frac{\pi}{2}} \,.
\end{gather}

In this article, we furthermore assume that the magnetic field $\vec{B}_0$ is stationary and homogeneous in the directions perpendicular to the probe photon propagation direction $\rm x$, i.e., $B=B({\rm x})$. Due to translational invariance in the perpendicular directions, there is no momentum transfer in these directions.
The integrations over the coordinates $\rm y$ and $\rm z$ in \Eqref{eq:Sp2} can then be performed right away.
In turn, the photons induced in the superposition of the pump and probe fields will exclusively propagate along $\rm x$. 
Limiting ourselves to terms linear in $b_0$, we obtain
\begin{multline}
{\cal S}_{(1)}(\vec{k})
 =\frac{i}{45}\frac{\alpha}{2\pi}\frac{b_0}{\sqrt{2\omega}}\biggl\{ (2\pi)^3\delta(|k_{\rm x}|-\omega)\delta(k_{\rm y})\delta(k_{\rm z}) \int{\rm dx}\,{\rm e}^{i(k_{\rm x}-\omega){\rm x}}\,\Bigl(\frac{eB({\rm x})}{m^2}\Bigr)^2  \\
  \times\Bigl[2(\omega-k_{\rm x}) \sin(\chi-\beta) 
   +4\omega\cos\beta\sin\chi+ 7\omega\sin\beta\cos\chi\Bigr] + {\cal O}\bigl((\tfrac{eB}{m^2})^4\bigr) + {\cal O}\bigl(\tfrac{eb_0}{m^2}\bigr){\cal O}\bigl(\tfrac{eB}{m^2}\bigr)
 \biggr\} \,,
 \label{eq:Sp2_2}
\end{multline}
where $\alpha=\frac{e^2}{4\pi}$.
In the conventional notion vacuum birefringence arises as an effect affecting photon propagation. In homogeneous fields it can, e.g., be derived straightforwardly from the photon polarization tensor evaluated in the given background field, generically mediating between an incident and an outgoing photon.
By means of the above linearizion in $b_0$ we specialize our calculation to the same effect, as we thereby limit ourselves to a single coupling to the incident probe photon field.
Generically, the above expression also accounts for contributions which are of higher-order in the coupling to the probe photon field, corresponding to photon merging effects, resulting in higher-harmonics.
Conversely, we have specialized to single-photon emission from the outset.
Finally, note that the same result should also be attainable by specializing the photon polarization tensor derived in Eq.~(10) of \cite{Karbstein:2015cpa} to weak magnetic fields and determining the photon current as in \cite{Karbstein:2015xra}.

The electric/magnetic peak field amplitude $b_0$ of the plane-wave probe can be related to the probe's mean intensity via $b_0=\sqrt{2\langle I\rangle}$.
The mean intensity in turn can be expressed as $\langle I\rangle =J\omega$,
where $J=\frac{N}{L_{\rm y}L_{\rm z}T}$ is the incident probe photon current, i.e., the number of probe photons per area $L_{\rm y}L_{\rm z}$ and time interval $T$.

Equation~\eqref{eq:Sp2_2} gives rise to two distinct induced photon contributions of energy $\omega$ propagating in positive and negative $\rm x$ direction, 
respectively.
Employing Fermi's golden rule it is convenient to define the numbers of induced frequency-$\omega$ photons in forward $(+)$ and backward $(-)$ directions as
\begin{equation}
 N^\pm_{(p)} = \int_{\mathbb{R}^\pm}\frac{{\rm d}k_{\rm x}}{2\pi}\int\frac{{\rm d}^2 k_\perp}{(2\pi)^2}\bigl|{\cal S}_{(p)}(\vec{k})\bigr|^2\,. \label{eq:Npm}
\end{equation}
Assuming the magnetic field amplitude to be even in $\rm x$, i.e., $B({\rm x})\equiv B(|{\rm x}|)$, and leaving the magnetic field profile unspecified for the moment,
the modulus squared of \Eqref{eq:Sp2_2} inserted into \Eqref{eq:Npm} becomes
\begin{multline}
 N^\pm_{(1)} 
 =N\frac{1}{45^2}\Bigl(\frac{\alpha}{2\pi}\Bigr)^2
 \biggl\{
 \biggl[\int{\rm dx}\,{\rm e}^{i(\omega\mp\omega){\rm x}}\,\Bigl(\frac{eB({\rm x})}{m^2}\Bigr)^2\biggr]^2 \\
  \times\Bigl[2(\omega\mp\omega) \sin(\chi-\beta) + 4\omega\cos\beta\sin\chi+ 7\omega\sin\beta\cos\chi\Bigr]^2 \\
   + {\cal O}\bigl((\tfrac{eB_0}{m^2})^6\bigr)
   + {\cal O}\bigl(\tfrac{eb_0}{m^2}\bigr){\cal O}\bigl((\tfrac{eB_0}{m^2})^3\bigr)
   + {\cal O}\bigl((\tfrac{eb_0}{m^2})^2\bigr){\cal O}\bigl((\tfrac{eB_0}{m^2})^2\bigr)
 \biggr\} \,,
 \label{eq:Sp2_3}
\end{multline}
for $p=1$ and $N^\pm_{(2)}=N^\pm_{(1)}|_{\chi\to\chi+\frac{\pi}{2}}$ for $p=2$,
where we made use of $\int\frac{{\rm d}^2k_\perp}{(2\pi)^2}(2\pi)^4\delta^2(k_{\rm y})\delta^2(k_{\rm z})=L_{\rm y}L_{\rm z}$, as well as $\int_{\mathbb{R}^\pm}\frac{{\rm d}k_{\rm x}}{2\pi}(2\pi)^2\delta^2(|k_{\rm x}|-\omega)f(k_{\rm x})=Tf(\pm\omega)$.

For the scenario considered here, we first span the two polarization modes of the induced photon signal
by the polarization vectors in Eqs.~\eqref{eq:epsilons}-\eqref{eq:epsilons1} specialized to $\vartheta=\frac{\pi}{2}$ and $\varphi=0$.
In turn, they can be expressed as $\epsilon^\mu_{(1)}(\hat{\vec{k}})=(0,\vec{e}_{\pi-\chi})$ and $\epsilon^\mu_{(2)}(\hat{\vec{k}})=\epsilon^\mu_{(1)}(\hat{\vec{k}})\big|_{\chi\to\chi+\frac{\pi}{2}}$.
Recalling that the electric field, and thus the polarization vector, of the incident probe wave points along $\hat{\vec{e}}=\vec{e}_{\beta+\frac{\pi}{2}}$, it is then convenient to decompose the induced photon signal into polarization components polarized parallel and perpendicular to the plane spanned by $\hat{\vec{k}}$ and $\hat{\vec{e}}$.
Hence, we finally define $\epsilon^\mu_{\parallel}(\hat{\vec{k}})\equiv\epsilon^\mu_{(1)}(\hat{\vec{k}})|_{\chi=\frac{\pi}{2}-\beta}=(0,\vec{e}_{\beta+\frac{\pi}{2}})$
and $\epsilon^\mu_{\perp}(\hat{\vec{k}})\equiv\epsilon^\mu_{(1)}(\hat{\vec{k}})|_{\chi=\pi-\beta}=(0,\vec{e}_\beta)$.
For the number of photons induced in these modes, we obtain
\begin{multline}
 N^\pm_\parallel
 =N\frac{1}{45^2}\Bigl(\frac{\alpha}{2\pi}\Bigr)^2
 \biggl\{
 \biggl[\int{\rm dx}\,{\rm e}^{i(\omega\mp\omega){\rm x}}\,\Bigl(\frac{eB({\rm x})}{m^2}\Bigr)^2\biggr]^2\omega^2
 \bigl[ (6\mp2)\cos^2\beta + (5\pm2)\sin^2\beta \bigr]^2 \\
   + {\cal O}\bigl((\tfrac{eB_0}{m^2})^6\bigr)
   + {\cal O}\bigl(\tfrac{eb_0}{m^2}\bigr){\cal O}\bigl((\tfrac{eB_0}{m^2})^3\bigr)
   + {\cal O}\bigl((\tfrac{eb_0}{m^2})^2\bigr){\cal O}\bigl((\tfrac{eB_0}{m^2})^2\bigr)
 \biggr\} \,, \label{eq:Npar}
\end{multline}
and
\begin{multline}
 N^\pm_{\perp}
 =N\frac{1}{45^2}\Bigl(\frac{\alpha}{2\pi}\Bigr)^2
 \biggl\{
 \biggl[\int{\rm dx}\,{\rm e}^{i(\omega\mp\omega){\rm x}}\,\Bigl(\frac{eB({\rm x})}{m^2}\Bigr)^2\biggr]^2 \omega^2
  \Bigl(\frac{4\mp1}{2}\Bigr)^2\sin^2(2\beta) \\
   + {\cal O}\bigl((\tfrac{eB_0}{m^2})^6\bigr)
   + {\cal O}\bigl(\tfrac{eb_0}{m^2}\bigr){\cal O}\bigl((\tfrac{eB_0}{m^2})^3\bigr)
   + {\cal O}\bigl((\tfrac{eb_0}{m^2})^2\bigr){\cal O}\bigl((\tfrac{eB_0}{m^2})^2\bigr)
 \biggr\} \,. \label{eq:Nperp}
\end{multline}
Clearly, the photons induced in forward direction give rise to the phenomenon of vacuum birefringence, as long as $\beta\neq n\frac{\pi}{2}$ with $n\in\mathbb{N}_0$:
The induced photons $N_\perp^{+}$ supplement the outgoing probe photon beam with photons polarized perpendicular to the incident probe beam, which can be interpreted as a birefringence signal \cite{Dinu:2013gaa,Karbstein:2015xra}.

The photons induced in backward direction constitute a quantum reflection signal \cite{Gies:2013yxa}, which is a direct consequence of the fact that an external electromagnetic field can be interpreted as constituting an attractive potential in the probe photons' equation of motion:
Probe photons originating from the zero-field region which are directed towards a magnetic field region of finite extent can experience above-barrier reflection \cite{QR1}.
Considering the same effect in an infinitely extended homogeneous background field, i.e., without imposing zero-field asymptotics for the incident probe as well as outgoing probe and signal photons, the probe does not sense any variation of the potential and no reflection signal is observed.

Note, that we obtained the same functional form for $N^{\pm}_\parallel$ already in \cite{Gies:2013yxa}. However, the approach employed there is not capable of unveiling the asymmetry in the coefficients of the forward and backward signals,
and thus results in the same numeric prefactors for both directions:
In \cite{Gies:2013yxa} we adapted the locally constant field approximation on the level of the photon polarization tensor.
Due to translational invariance, the polarization tensor in homogeneous fields does only depend on a single four momentum $k^\mu$.
However, the reflection process $(-)$ under consideration genuinely mediates momenta $k^\mu=(k^0,k_{\rm x},0,0)$ into $k'^\mu=(k^0,-k_{\rm x},0,0)$, while $k'^\mu=k^\mu$ in forward direction $(+)$.
In turn, terms of the type $k_\mu k'^\mu=k_{\rm x}^2\mp(k^0)$ distinguishing between the $\pm$ directions are not accounted for in \cite{Gies:2013yxa}.
Similar modifications are to be expected for the other scenarios \cite{Gies:2014wsa} (based on the approach devised in \cite{Gies:2013yxa}) for which quantum reflection has been investigated so far.

\begin{figure}
 \centering
  \includegraphics[width=0.6\textwidth]{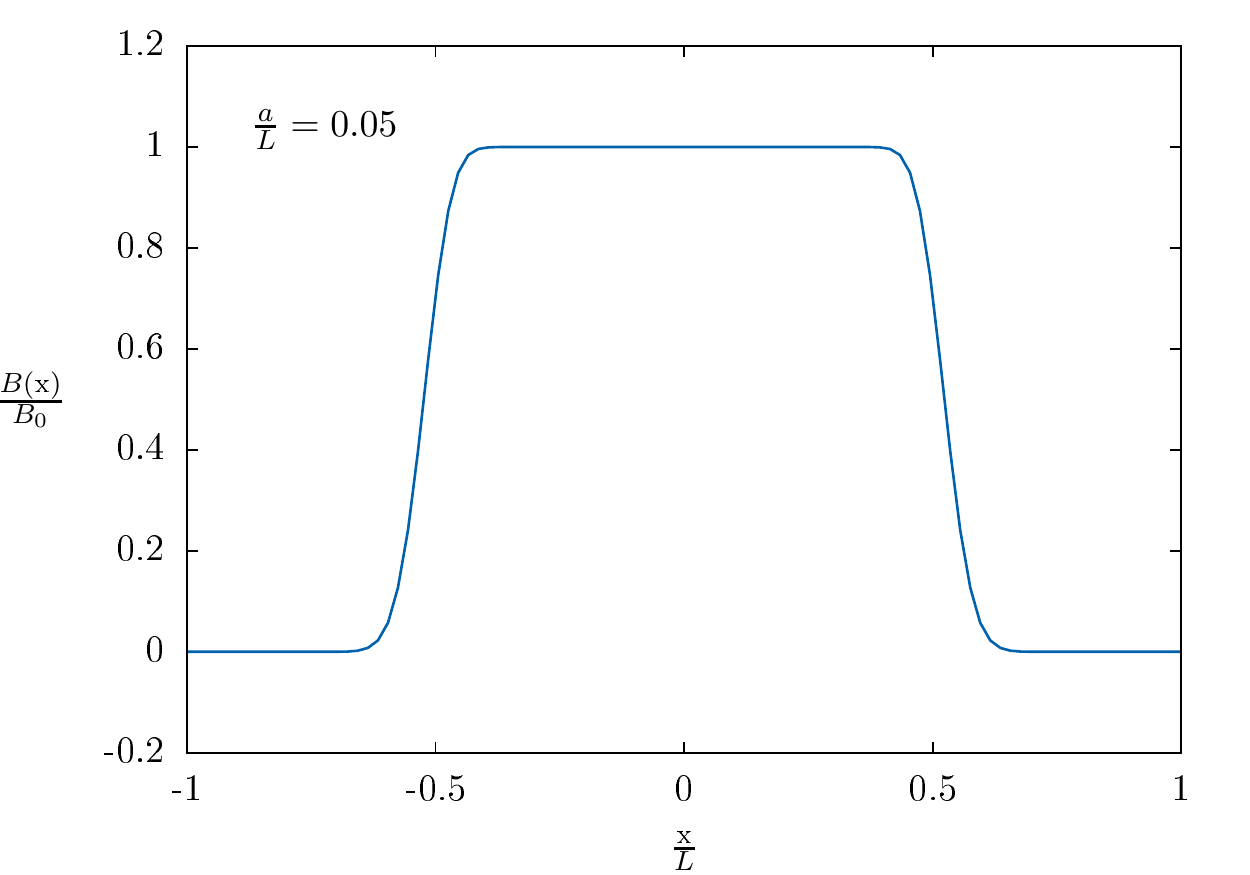} 
\caption{Exemplary plot of the magnetic field amplitude profile~\eqref{eq:Bprofile}.}
\label{fig:feld}
\end{figure}

Let us finally adopt the following specific magnetic field amplitude profile,
\begin{equation}
 B({\rm x})=\frac{1}{\sqrt{2}}B_0\Bigl[{\rm erf}\Bigl(\tfrac{1}{a}\bigl({\rm x}+\tfrac{L}{2}\bigr)\Bigr)-{\rm erf}\Bigl(\tfrac{1}{a}\bigl({\rm x}-\tfrac{L}{2}\bigr)\Bigr)\Bigr]^{1/2}, \label{eq:Bprofile}
\end{equation}
depicted in Fig.~\ref{fig:feld}, which can be straightforwardly Fourier transformed to momentum space,
$\int{\rm dx}\,{\rm e}^{i\kappa{\rm x}}B^2({\rm x})=\frac{2}{\kappa}B_0^2\sin\bigl(\frac{\kappa}{2}L\bigr)\,{\rm e}^{-(\frac{\kappa}{2}a)^2}$.
Obviously, this magnetic field profile fulfills $\int{\rm dx}\,B^2({\rm x})=LB_0^2$, which is completely independent of the additional length scale $a$.
Hence, we obtain the same expressions for $N^+_{\parallel/\perp}$ as if we had naively adapted the homogeneous field result to a finite length interval, i.e., assumed a constant field amplitude $B_0$ extending over a length $L$.
The second length scale $a$ governs the distance over which the field amplitude increases from $0$ to $B_0$, and analogously drops from $B_0$ to $0$.
This becomes most transparent on the level of $B^2({\rm x})$, by noting that $\frac{\rm d}{{\rm dx}}\,{\rm erf}\bigl(\tfrac{1}{a}({\rm x}\pm\tfrac{L}{2})\bigr)=\frac{2}{\sqrt{\pi}}\frac{1}{a}\,{\rm e}^{-\frac{1}{a^2}({\rm x}\pm\frac{L}{2})^2}$.
In order not to leave the range of applicability of the locally constant field approximation, we have to ensure that $a\gg\lambda_C$.
Upon insertion of this magnetic field amplitude profile into Eqs.~\eqref{eq:Npar}-\eqref{eq:Nperp} and choosing $\beta=\frac{\pi}{4}$, which maximizes the birefringence signal $N^\pm_{\perp}$, we obtain
\begin{multline}
\left\{\begin{array}{c}
 N^+_\parallel \\ N^+_{\perp}
\end{array}\right\}
 =N\frac{1}{45^2}\Bigl(\frac{\alpha}{4\pi}\Bigr)^2
 \biggl\{
 \Bigl(\frac{eB_0}{m^2}\Bigr)^4(\omega L)^2
 \left\{\begin{array}{c}
 11^2 \\ 3^2
\end{array}\right\}
 \\
   + {\cal O}\bigl((\tfrac{eB_0}{m^2})^6\bigr)
   + {\cal O}\bigl(\tfrac{eb_0}{m^2}\bigr){\cal O}\bigl((\tfrac{eB_0}{m^2})^3\bigr)
   + {\cal O}\bigl((\tfrac{eb_0}{m^2})^2\bigr){\cal O}\bigl((\tfrac{eB_0}{m^2})^2\bigr)
 \biggr\} \,,
\end{multline}
and
\begin{multline}
\left\{\begin{array}{c}
 N^-_\parallel \\ N^-_{\perp}
\end{array}\right\}
 =N\frac{1}{45^2}\Bigl(\frac{\alpha}{4\pi}\Bigr)^2
 \biggl\{
  \Bigl(\frac{eB_0}{m^2}\Bigr)^4\sin^2(\omega L)\,{\rm e}^{-2(\omega a)^2}
 \left\{\begin{array}{c}
 11^2 \\ 5^2
\end{array}\right\}\\
   + {\cal O}\bigl((\tfrac{eB_0}{m^2})^6\bigr)
   + {\cal O}\bigl(\tfrac{eb_0}{m^2}\bigr){\cal O}\bigl((\tfrac{eB_0}{m^2})^3\bigr)
   + {\cal O}\bigl((\tfrac{eb_0}{m^2})^2\bigr){\cal O}\bigl((\tfrac{eB_0}{m^2})^2\bigr)
 \biggr\} \,.
\end{multline}
for the induced photon numbers in forward and backward direction, respectively.
The result for $N^+_{\perp}$ agrees with the one obtained by the conventional calculation, assuming the probe photons to traverse a constant magnetic field $B_0$ of length $L$:
Here, the number of probe photons scattered in the perpendicular polarization mode is obtained from the phase shift $\Delta\phi=\omega L\Delta n$, with $\Delta n=\frac{\alpha}{4\pi}\frac{6}{45}\bigl(\frac{eB_0}{m^2}\bigr)^2$ \cite{Toll:1952},
as $N_\perp=N(\frac{\Delta\phi}{2})^2$.

Note that the numbers of photons induced in backward direction constituting the quantum reflection signal are generically suppressed by an overall factor of $\bigl(\frac{\sin(\omega L)}{\omega L}{\rm e}^{-(\omega a)^2}\bigr)^2$
in comparison to the numbers of photons induced in forward direction.

\section{Conclusions}

In this article we have exemplarily studied vacuum magnetic birefringence as a vacuum emission process.
However, our approach is not limited to the magnetic field case but also applicable to other and more sophisticated field configurations, the only restriction being the limitation to slowly varying electromagnetic fields.
Favorably, many field configurations available in the laboratory fall into this class.
Insights beyond the locally constant field approximation could, e.g., be obtained from an {\it ab initio} worldline numeric evaluation of the photon polarization tensor in a given field inhomogeneity \cite{Gies:2011he}.
We have shown that photons originating at zero-field directed towards a localized strong-field region experience the effect of quantum reflection off the strong-field.
Due to translational invariance in infinitely extended constant backgrounds, this phenomenon is not encountered in such backgrounds.

\ack

I am grateful to the organizers of PHOTON 2015 for organizing a very nice and stimulating conference.
Moreover, I would like to thank H.~Gies, M.~Reuter, N.~Seegert, R.~Shaisultanov, and M.~Zepf for collaboration on related topics, B.~D\"obrich and G.~Zavattini for stimulating discussions during the conference, and V.~Serbo for bringing Ref.~\cite{Kotkin:1996nf} to my attention.
Finally, I am indebted to B.~D\"obrich and H.~Gies for useful comments on this manuscript.

\end{document}